# Near Infrared polarimetry of a sample of YSOs*

A. Pereyra[1,2], J. M. Girart[3], A. M. Magalhães[2], C. V. Rodrigues[4], and F. X. de Araújo[1]

[1] Observatório Nacional, Rua General José Cristino 77, São Cristovão, 20921-400, Rio de Janeiro, Brazil
[2] Departamento de Astronomia, IAG, Universidade de São Paulo, Rua do Matão 1226, São Paulo, SP, 05508-900, Brazil
[3] Institut de Ciències de l'Espai (CSIC- IEEC), Campus UAB - Facultat de Ciències, Torre C5 - parell 2, 08193 Bellaterra, Catalunya, Spain
[4] Instituto Nacional de Pesquisas Espaciais/MCT, Avenida dos Astronautas 1758, São José dos Campos, SP, 12227-010, Brazil

**ABSTRACT**

*Aims.* Our goal is to study the physical properties of the circumstellar environment of young stellar objetcs (YSOs). In particular, the determination of the scattering mechanism can help to constrain the optical depth of the disk and/or envelope in the near infrared.
*Methods.* We used the IAGPOL imaging polarimeter along with the CamIV infrared camera at the LNA observatory to obtain near infrared polarimetry measurements at the $H$ band of a sample of optically visible YSOs, namely, eleven T Tauri stars and eight Herbig Ae/Be stars. An independent determination of the disk (or jet) orientation was obtained for twelve objects from the literature. The circumstellar optical depth could be then estimated comparing the integrated polarization position angle (PA) with the direction of the major axis of the disk projected in the plane of the sky. In general, optically thin disks have polarization PA perpendicular to the disk plane. In contrast, optically thick disks produce polarization PA parallel to the disks.
*Results.* Among the T Tauri stars, three are consistent with optically thin disks (AS 353A, RY Tau and UY Aur) and five with optically thick disks (V536 Aql, DG Tau, DO Tau, HL Tau and LkH$\alpha$ 358). Among the Herbig Ae/Be stars, two stars show evidence of optically thin disk (Hen 3-1191 and VV Ser) and two of optically thick disks (PDS 453 and MWC 297). Our results seem consistent with the fact that optically thick disks at near infrared bands are associated more likely with younger YSOs. Marginal evidence of polarization reversal is found in RY Tau, RY Ori, WW Vul, and UY Aur. On the first three cases this feature can be associated to the UXOR phenomenon. Correlations with the *IRAS* colours and the spectral index yielded evidence of an evolutionary segregation with the disks tend to be optically thin when they are older.

## 1. Introduction

The study of the circumstellar environment in young stellar objects (YSOs) is one of the most interesting topics in star formation and can give clues about the birth of planetary systems. Polarimetry is a powerful tool to detect asymmetries in YSOs. This is specially important for non-resolved objects where the technique gives a direct diagnostic. The scattered emission produced by gas or dust in outflows, jets and/or disks produces polarization that can give us information about the physical properties of these environments.

In general, polarimetric observations in YSOs show temporal variability usually with the polarization position angle (PA) confined to a narrow interval that allows to infer the presence of a disk (Bastien 1996). Early studies in YSOs comparing the PA of the linear polarization and the direction of the jets indicated that they are preferentially perpendicular (Mundt & Fried 1983). Single scattering in an optically thin disk (Brown & McLean 1977) results in a PA perpendicular to the direction of the major axis of the disk projected in the plane of the sky (hereinafter, the disk PA).

On the other hand, optically thick disks may show multiple scattering with the PA parallel to the disk PA (Angel 1969; Bastien & Ménard 1990). In general, the effect of different disk inclinations alters the polarization level but maintains the PA unchanged in both scenarios. For pole-on objects, no net polarization is observed and there is no sense in defining a PA for these systems.

Alternative models (Whitney & Hartmann 1992, 1993) also can obtain polarization PA perpendicular to the disk PA in geometrically thin but optically thick disks. **These models can also generate polarization PA parallel to the disk PA if there is scattering by material above and below the equatorial disk.** In these cases, the optically thick disk requires an additional envelope with bipolar cavities cleared by an outflow from the YSOs.

The statistics from Ménard & Bastien (1992) is compatible with the prevalent presence of optically thick circumstellar disks surrounding YSOs. Their sample included mainly T Tauri stars (TTS) but also some Herbig Ae/Be stars (HeAeBeS). Maheswar et al. (2002) compiled a sample of HeAeBeS with outflows direction well defined along with polarization measurements and concluded that 55% of them are consistent with optically thick disks. This seems to indicate that disks in YSOs are usually optically thick. This

---

is also supported by Bastien (1996) who concluded that the polarization observations in YSOs can be explained by multiple scattering by dust grains in a flattened circumstellar disk probably embedded in larger, approximately spherical envelope.

Polarimetry in the near infrared (NIR) is less affected by foreground polarization, something that can be critical in optical bands. Correlations between the NIR polarization PA and any existing asymmetry (like outflows, jets or disks) allow thus a better evaluation of the optical depth of the scattering mechanism. Compared with the optical, the statistics for NIR polarization measurements of YSOs needs to be improved (Hough et al. 1981; Moneti et al. 1984; Tamura & Sato 1989; McGregor et al. 1994; Whitney et al. 1997; Perrin et al. 2004; Beckford et al. 2008; Kusakabe et al. 2008). Therefore, a larger NIR sample will help to clarify if the trend of optically thick disks in optical wavelengths is also valid in the NIR domain. In addition, the comparison between the optical and NIR polarization PA can give clues about the detection of a possible polarization PA reversal in YSOs (Hough et al. 1981), caused by the transition from an optically thick disk in the optical to an optically thin environment in the NIR.

Asymmetries in YSOs can also be explored using radio, mm/submm data, or NIR interferometry. For example, the disk PA can be evaluated using CO emission (Korner & Sargent 1995) or thermal continuum emission from dust (Kitamura et al. 2002; Qi et al. 2003). NIR interferometry is also useful to detect the disk PA in YSOs (Akeson et al. 2005; Malbet et al. 2007). On the other hand, the PA of outflows or jets can be obtained from radio extended emission (Anglada 1996). In addition, long-slit spectroscopy in forbidden emission lines has also been used to obtain the jet axis in YSOs (Hirth et al. 1994, 1997; Lavalley et al. 1997; Mund & Eislöffel 1998).

In this work, we combine the detection of asymmetries in YSOs through NIR polarization with the determination of the disk PA from other techniques to constrain the optical properties of the circumstellar disks. Our sample includes selected TTS and HeAeBeS. The observations and data reduction are presented in § 2. The results, including the foreground and intrinsic polarization computation, are presented in § 3. The discussion about the correlations with the jet/disk PA and the comparison between optical and NIR polarization PA are shown in § 4. This section also includes correlations between the disk optical depth and the *IRAS* colours and the spectral index. Comments on individual objects are presented in § 5. The conclusions are drawn in § 6.

## 2. Observations

The observations were made using IAGPOL, the IAG imaging polarimeter (Magalhães et al. 1996), at the Observatório do Pico dos Dias (OPD), Brazil. Additional details about this instrument can be found in Pereyra (2000) and Pereyra & Magalhães (2002).

The 0.6m and 1.6m LNA telescopes were used in several runs between 2003 and 2005. The CamIV infrared camera with an $H$ broadband filter was used along with IAGPOL. CamIV is based on a HAWAII detector (manufactured by Rockwell Sci.) of 1024×1024 pixels and 18.5 $\mu$m/pixel that yields a plate scale of 0$\farcs$5/pixel and 0$\farcs$25/pixel at the 0.6m and 1.6m telescopes, respectively. The retarders used by IAGPOL were an achromatic $\lambda/2$-waveplate centered in 800nm (manufactured by Meadowlark Optics, Inc.) and an achromatic $\lambda/4$-waveplate optimized to NIR (manufactured by B. Halle Nachfl[1]). Each object was observed through eight waveplate positions (WP) separated by 22$\fdg$5 with five dithered images *per* WP. A total of (8×5=) 40 images were gathered *per* object. Density filters were used when appropriate.

At each WP the sky was computed using the mode of the dithered positions. After sky subtraction, flat correction and registration of the images, the PCCDPACK package (Pereyra 2000) was used to compute the polarimetric parameters. The CamIV's detector yielded a typical area of 8′ and 4′ for the 0.6 and 1.6m telescopes, respectively. A log of observation is shown in Table 1 along with optical and NIR magnitudes and extinction estimates for each object.

Observations of unpolarized stars through a Glan prism yielded the effective retardation ($\tau$), the fiducial zero point of the system (zero), and the polarimetric efficiency of the instrument in each run. The efficiency was always close to 100% for each ($\tau$, zero) solution and no efficiency correction was applied to the data. Observations of polarized standard stars in each run yielded the position angle correction to the equatorial system (PA$_{\rm corr}$). Unpolarized standard stars were used to check the instrumental polarization. This was found to be less than 0.3%; no such correction was then applied to the data. A summary of the calibration data for each run is shown in Table 2.

The selection criteria of our sample was YSOs optically visible and with measured optical polarization (at least in one epoch) higher than 1% (see Table 3). A summary of the temporal evolution of the optical polarization (mainly at $V$ band) for the objects of our sample is also included in Table 3. The observation epoch for the optical polarization data is also indicated along with the number of measurements *per* epoch ($N$), with mean values quoted.

## 3. Results

### 3.1. Observed Polarization

The observed polarization for the objects of our sample is shown in Table 4. Column (2) shows the aperture radius (in arcsecs) of the observations. These are computed for a serie of apertures around the object; we select the one that minimized the polarization accuracy. This accuracy is calculated from the spread of the measurements at each WP with regards to the 4-cosine curve. Columns (3) and (4) show the observed polarization in $H$ band along with its PA, respectively. As we can note, the NIR polarization level is higher than 1% for all cases except in three T Tauri stars (AS 352A, RY Tau, and UY Aur) and one Herbig Ae/Be (MWC 297). Rather high observed polarization levels (higher than 4%) are found in three TTS (V536 Aql, HL Tau and LkH$\alpha$ 358) and in one HeAeBeS (PDS 406).

Only three TTS (RY Tau, DG Tau, and HL Tau) of our sample have a previous NIR polarization measurement in $H$ band (see Table 7). For the remaining objects we present their first NIR polarization data, to the best of our knowledge.

---
[1] http://www.b-halle.de

**Table 1.** Log of observations.

| Object | Other name | $V^a$(mag) | $H^b$(mag) | $A_V^c$(mag) | Date | Telescope | Total IT$^d$ (s) |
|---|---|---|---|---|---|---|---|
| | | | | T Tauri | | | |
| V895 Sco | Haro 1-1 | 13.3 | 9.3 | 1.4 | 2005/jun/25 | 1.6m | 60 |
| AS 353A | V1352 Aql | 12.5 | 9.2 | 5.7 | 2005/jun/25 | 1.6m | 60 |
| PX Vul | Hen 3-1751 | 11.7 | 8.5 | 1.8 | 2005/jun/25 | 1.6m | 60 |
| V536 Aql | | 12.6 | 8.1 | 1.0 | 2005/jun/24 | 1.6m | 60 |
| RY Tau | HD 283571 | 10.2 | 6.1 | 0.5 | 2005/oct/12 | 0.6m | 32 |
| DG Tau | | 12.8 | 7.7 | 3.3 | 2005/oct/12 | 0.6m | 120 |
| DO Tau | | 13.5 | 8.2 | 4.1 | 2005/oct/13 | 0.6m | 200 |
| HL Tau | Haro 6-14 | 14.6 | 9.2 | 2.2 | 2005/oct/12 | 0.6m | 400 |
| LKH$\alpha$ 358$^e$ | | 19.0 | 10.9 | 2.2 | 2005/oct/12 | 0.6m | 400 |
| UY Aur | | 12.4 | 8.0 | 1.9 | 2005/oct/13 | 0.6m | 140 |
| RY Ori | Haro 5-82 | 11.8 | 8.9 | 0.6 | 2003/sep/20 | 0.6m | 600 |
| | | | | Herbig Ae/Be | | | |
| PDS 406 | | 13.9 | 10.9 | 0.8 | 2005/jun/24 | 1.6m | 600 |
| Hen 3-1191 | | 13.7 | 8.1 | 1.6 | 2005/jun/24 | 1.6m | 40 |
| | | | | | 2005/jun/26 | 1.6m | 40 |
| PDS 453 | | 12.9 | 10.0 | 1.1 | 2005/jun/24 | 1.6m | 120 |
| MWC 297 | PDS 518 | 12.3 | 4.4 | 5.4 | 2005/oct/13 | 0.6m | 44$^f$ |
| VV Ser | | 12.2 | 7.4 | 5.7 | 2005/jun/26 | 1.6m | 32 |
| PDS 520 | | 14.7 | 8.6 | 5.5 | 2005/jun/25 | 1.6m | 40 |
| WW Vul | HD 344361 | 10.5 | 8.2 | 0.4 | 2005/jun/26 | 1.6m | 40 |
| VY Mon | | 12.9 | 6.7 | 2.6 | 2005/oct/13 | 0.6m | 40 |

(a) from SIMBAD, except for: RY Ori and VV Ser (from NOMAD catalog, Zacharias et al. 2004), and PDS 520 (from Vieira et al. 2003).
(b) 2MASS magnitude at H band.
(c) from Dobashi et al. (2005)
(d) total integration time at eight waveplate positions, including five ditherings *per* position.
(e) object in the same field of HL Tau.
(f) with density filter DN2.6

### 3.2. Foreground Polarization

In order to estimate the NIR foreground polarization toward each target, we searched the literature for previous such estimates in the optical. Four objects have the optical foreground polarization already computed: AS 353A (Monin et al. 2006), RY Tau (Petrov et al. 1999), MWC 297 (Hillenbrand et al. 1992), and WW Vul (Grinin et al. 1988). Additional optical foreground polarization for seven objects of our sample was obtained from a polarimetric survey of HeAeBeS (Rodrigues et al. 2009). In this survey, the foreground polarization was estimated by the weighted average of the polarization of field stars within typically $10' \times 10'$ around the objects. This foreground estimate was choosen when more than one computation was available (as in MWC 297 and WW Vul).

We assumed that the foreground polarization can be represented by a standard Serkowski law, $P = P_{max} \exp[-K\ln^2(\lambda_{max}/\lambda)]$ (Serkowski et al. 1975), with the maximum polarization, $P_{max}$, assumed to be at $\lambda_{max} = 0.55\mu m$ and the parameter $K$ depending on $\lambda_{max}$ following the relationship of Whittet et al. (1992). Then, we extrapolated the Serkowski relation to the NIR wavelength and obtained the proper foreground polarization for $H$ band (1.65$\mu$m). We have assumed that the foreground PA is constant with wavelength. The foreground polarization computed in this way is 1/3 of the $P_{max}$ assumed and is shown in columns (5) and (6) of Table 4. Column (7) indicates the number of the field objects used to compute the foreground polarization by Rodrigues et al. (2009). In general, a better statistic ($N$) will yield a more precise estimator.

For the nine objects whose foreground correction is available, just two (RY Tau and MWC 297) appear with a foreground polarization level higher than the observed one. Nevertheless, in all the cases, the foreground correction is lower than 1% and as expected we have a substantial reduction in the foreground at NIR bands compared to the optical polarization. To help the comparison, Figure 1 shows for each object the Stokes parameters[2] $Q-U$ diagram, including the NIR observed polarization and the optical and NIR vectors for the foreground polarization (when available). Analyzing these diagrams, it seems clear that the contribution of the foreground polarization to the observed polarization at NIR bands is significantly reduced.

### 3.3. Intrinsic polarization

For those objects with an estimated foreground polarization (Table 3), we obtained the intrinsic polarization. As the foreground polarization is an additive component included in the observed polarization, the intrinsic Stokes parameters are as follows:

$$Q_{int} = Q_{obs} - Q_{for}$$
$$U_{int} = U_{obs} - U_{for}.$$

The intrinsic polarization ($P_{int}$) and its polarization angle (PA$_{int}$) are obtained from:

$$P_{int} = (Q_{int}^2 + U_{int}^2)^{1/2}$$
$$PA_{int} = \tfrac{1}{2}\arctan(U_{int}/Q_{int}).$$

---
[2] $Q = P\cos(2\times PA)$ and $U = P\sin(2\times PA)$

The intrinsic polarization estimated for the nine objects (two TTS and seven HeAeBeS) are indicated in columns (8) and (9) of Table 4. In all cases, the intrinsic polarization levels are significant (higher than 0.9%). In seven objects the difference between the observed and intrinsic polarization levels is less or equal to 0.3%. Only RY Tau and PDS 406 show larger polarization changes after the foreground correction (0.9% and 0.6%, respectively). In addition, seven YSOs show small changes in the PA when correcting for foreground polarization, typically similar or less than 10°. Significant deviations are found in PDS 453 (19°) and MWC 297 (41°).

As we can note, in general, the subtraction of the foreground polarization does not change significatively the results. For example, the extreme $P_{int}$ measured for PDS 406 (4.2%) is comparable with its $P_{obs}$ (4.8%) with the PA practically unchanged. We can then confirm the small contribution of the foreground polarization. This is important if the foreground polarization is not available and only $PA_{obs}$ can be used to infer correlations with the position angle for the jets/disks.

## 4. Discussion

### 4.1. Correlations with position angles of jets and disks

The main focus of this work is to correlate the polarization PA with the directions of jets, outflows or disks/envelope for the objects of our sample available in the literature. We found non-polarimetric evidence of asymmetries in twelve objects of our sample (eight TTS and four HeAeBeS). The PA for disks or jets ($PA_{jet}$ and $PA_{disk}$) are indicated in Table 5 with their respective references.

In five cases, there are independent determinations of $PA_{jet}$ and $PA_{disk}$ and within the uncertainties they are perpendicular. This is quantified by the parameter $\Delta_{jet/disk}$ in Table 5. For RY Tau, DG Tau, and DO Tau, $PA_{jet}$ and $PA_{disk}$ are perpendicular within 4° and for HL Tau and MWC 297 within 26°.

The angle difference between the polarization PA and the disk PA is defined as $\Delta_{pol/disk}$ in Table 5. For an optically thick disk, this difference is 0°, whereas for an optically thin disk the difference is 90°. We have classified the disk optical depth preferentially using the intrinsic polarization PA. If it was not available the observed one was used.

In the twelve objects where the comparison can be done, the jet/disk PA is at least within 14° of being parallel or perpendicular to the polarization PA. HL Tau and MWC 297, the two YSOs with discrepancies between the jet and disk PA, are better correlated with the disk PA. In summary, seven objects present optically thick disk and five optically thin disks. This is shown with a respective label at Table 5. Our classification is also indicated in the histogram of $\Delta_{pol/disk}$ (Fig. 2). Individual comments for each analysed object are presented in § 5.

It is interesting to note that the optically thick disks of our sample present the highest observed polarizations (Fig. 3). This result is consistent with the expected enhanced polarization by multiple scattering in an optically thick environment (Ménard & Bastien 1992). Alternatively, these polarization levels also can be explained by less dilution of the observed radiation by direct light of the central object in an optically thick line of sight (as in UXOR stars - see next section).

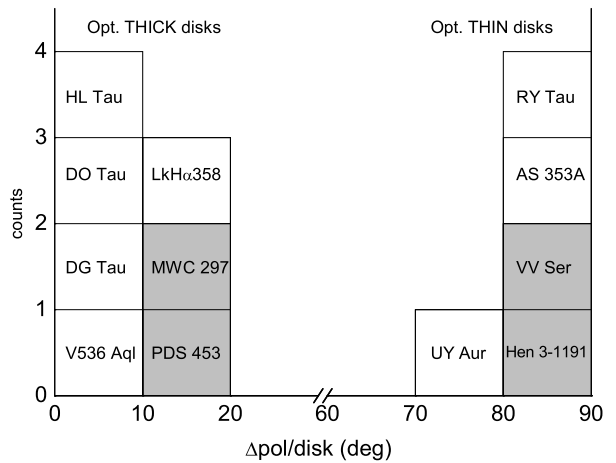

**Fig. 2.** Histogram of the $\Delta_{pol/disk}$ parameter used to classify the disk optical depth at NIR bands. The TTS are in white boxes and the HeAeBeS in grey boxes.

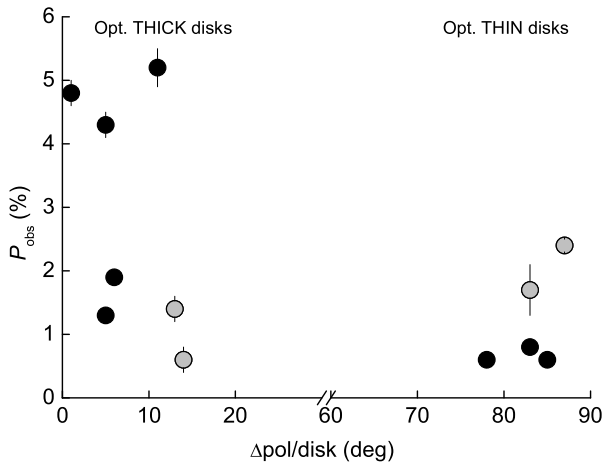

**Fig. 3.** Correlation between the observed polarization at NIR bands and the $\Delta_{pol/disk}$ parameter. The TTS are in black dots and the HeAeBeS in grey dots.

### 4.2. Comparing optical and NIR polarization PA - cases of possible reversal?

A change in the polarization PA by 90° when going from the visible to the NIR (or polarization reversal) has been observed in a couple of YSOs (T Tau and SU Aur, Hough et al. 1981). This feature can be explained by an optically thick disk at the optical bands, producing a polarization parallel to the disk, but optically thin at NIR bands, resulting in a polarization PA perpendicular to the disk (Bastien & Ménard 1990).

Evidence of polarization reversal in our sample can be obtained comparing the optical and NIR polarization PA from Tables 3 and 4. A direct comparison can be done analizing the $Q-U$ diagrams shown in Figure 1. The polarization reversal will be evident when a flip between two opposite quadrants in the the $Q-U$ diagram exists between the optical and NIR polarization. As the comparison is done

**Table 2.** Calibration summary.

| Run | Telescope | Retarder | $\tau$ (°) | zero (°) | Efficiency (%) | PA$_{\text{corr}}$ (°) |
|---|---|---|---|---|---|---|
| 2003/sep | 0.6m | $\lambda/2-800$nm | 141 | 35 | 100.1 (0.2)[a] | +138.0 |
| 2004/may | 1.6m | $\lambda/2-800$nm | 141 | 35 | 98.4 (2.7) | +59.3 |
| 2005/jun | 1.6m | $\lambda/4-$NIR | 87 | 47 | 99.4 (0.4)[b] | +51.2 |
| 2005/oct | 0.6m | $\lambda/4-$NIR | 87 | 47 | 99.4 (1.3) | +41.9 |

errors in parenthesis.
(a) mean of three measurements.
(b) mean of two measurements.

**Table 3.** Summary of the temporal evolution of the optical polarization of the targets from the literature.

| Object | date/epoch | $N$ | band | $P^a$ (%) | PA$^a$ (°) | ref. |
|---|---|---|---|---|---|---|
| | | T Tauri | | | | |
| V895 Sco | 1978/05 | 1 | $V$ | 3.0 (0.4) | 165 | 1 |
| | 1978/05 | 1 | $R$ | 2.9 (0.2) | 173 | 1 |
| AS 353A | 1976/07−1978/07 | 7 | $V$ | 1.4 (0.1) | 145 | 1 |
| | 1976/10−1978/07 | 6 | $R$ | 1.3 (0.02) | 144 | 1 |
| | 1986/08 | 1 | $R$ | 1.4 (0.2) | 139 | 4 |
| | 2000/05 | 1 | $I$ | 1.4 (0.04) | 125 | 5 |
| PX Vul | 1998/07−1998/10 | 9 | $V$ | 3.9 (0.1) | 28 | 2 |
| V536 Aql | 1977/08−1978/07 | 2 | $V$ | 6.6 (0.3) | 67 | 1 |
| | 1977/08−1978/07 | 2 | $R$ | 5.5 (0.5) | 65 | 1 |
| | 1985/06−1986/08 | 2 | $R$ | 7.6 (0.4) | 43 | 4 |
| RY Tau | 1972/01 | 1 | $V$ | 3.2 (0.2) | 30 | 9 |
| | 1973/01 | 1 | $V$ | 5.0 (0.1) | 20 | 9 |
| | 1976/09−1978/12 | 16 | $V$ | 2.5 (0.2) | 21 | 1 |
| | 1979/09 | 1 | $V$ | 3.1 (0.2) | 18 | 8 |
| | 1980/01 | 3 | $V$ | 3.0 (0.04) | 18 | 6 |
| | 1980/02 | 1 | $V$ | 2.2 (0.1) | 39 | 8 |
| | 1998/10−1999/01 | 7 | $V$ | 2.6 (0.1) | 23 | 2 |
| | 2001/12−2003/12 | 4 | $R$ | 2.0 (0.3) | 13 | 10 |
| DG Tau | 1976/10−1978/10 | 3 | $V$ | 5.5 (0.1) | 136 | 1 |
| | 1976/10−1978/10 | 2 | $R$ | 6.0 (0.1) | 135 | 1 |
| | 1980/01 | 1 | $V$ | 5.1 (0.1) | 137 | 6 |
| DO Tau | 1976/10 | 1 | $V$ | 3.1 (0.2) | 171 | 1 |
| | 1976/10 | 1 | $R$ | 2.9 (0.1) | 171 | 1 |
| | 1980/01 | 1 | $V$ | 3.8 (0.3) | 175 | 6 |
| HL Tau | 1976/10 | 1 | $V$ | 8.3 (0.4) | 146 | 1 |
| | 1976/10 | 1 | $R$ | 11.2 (0.2) | 147 | 1 |
| | 1980/01 | 1 | $V$ | 13.5 (0.8) | 143 | 6 |
| | 1985/01 | 1 | $B$ | 13.7 (0.3) | 138 | 4 |
| | 1987/01 | 1 | $clear$ | 14.0 (0.3) | 143 | 11 |
| LkH$\alpha$ 358 | 1987/01 | 1 | $clear$ | 4.1 (0.3) | 6 | 11 |
| UY Aur | 1978/12/01 | 1 | $V$ | 3.3 (0.2) | 127 | 1 |
| | 1979/02/28 | 1 | $V$ | 2.3 (0.5) | 162 | 1 |
| | 1980/01/05 | 1 | $V$ | 3.0 (0.2) | 161 | 6 |
| RY Ori | 1995/01/30 | 1 | $V$ | 6.6 (1.9) | 165 | 7 |
| | 1995/01/30 | 1 | $V$ | 4.3 (1.9) | 41 | 7 |
| | 1998/10−1999/02 | 7 | $V$ | 2.7 (0.1) | 72 | 2 |
| | | Herbig Ae/Be | | | | |
| PDS 406 | 1999/04/11 | 1 | $V$ | 4.7 (0.1) | 34 | 3 |
| Hen 3-1191 | 1999/04/11 | 1 | $V$ | 5.9 (0.2) | 46 | 3 |
| PDS 453 | 1999/04/08 | 1 | $V$ | 3.6 (0.1) | 49 | 3 |
| MWC 297 | 1973/10 | 1 | $V$ | 1.4 (0.2) | 95 | 12 |
| | 1975/08 | 1 | $V$ | 1.9 (0.1) | 98 | 12 |
| | 1976/06 | 1 | $V$ | 1.8 (0.1) | 95 | 12 |
| | 1985(6) | 1 | $V$ | 2.1 (0.2) | 91 | 13 |
| | 1998/05 | 1 | $V$ | 1.2 (0.2) | 95 | 2 |
| | 1999/07/29 | 1 | $V$ | 1.8 (0.1) | 94 | 3 |
| VV Ser | 1998/05−1998/10 | 11 | $V$ | 1.7 (0.1) | 78 | 2 |
| | 2000/06/22 | 1 | $V$ | 1.8 (0.1) | 78 | 3 |
| PDS 520 | 1999/07/28 | 1 | $V$ | 3.5 (0.1) | 15 | 3 |
| WW Vul | 1987/08 | 3 | $V$ | 5.3 (0.6) | 170 | 14 |
| | 1998/05 | 3 | $V$ | 0.6 (0.1) | 93 | 2 |
| | 1998/07−1998/10 | 8 | $V$ | 0.4 (0.1) | 146 | 2 |
| | 2000/06/21 | 1 | $V$ | 0.9 (0.1) | 151 | 3 |
| VY Mon | 1995/01−1995/02 | 7 | $V$ | 8.7 (0.7) | 6 | 7 |
| | 1998/10−1999/02 | 3 | $V$ | 9.8 (0.5) | 5 | 2 |

errors in parenthesis.
[a] mean values quoted when $N > 1$.
References: (1) Bastien (1982), the $\sigma_1$ error is quoted for individual observations and $\sigma_2$ for mean values; for simplicity $V$ and $R$ bands refer to the filters centered in 5895Å and 7543Å, respectively; only $V$ measurements for RY Tau and UY Aur are quoted; (2) Oudmaijer et al. (2001), only $V$ measurements are quoted; (3) Rodrigues et al. (2009); (4) Ménard & Bastien (1992); (5) Monin et al. (2006); (6) Bastien (1985), for simplicity $B$ refers to the filter centered in 4700Å and $R$ to the filters centered in 7640Å and 7675Å; (7) Yudin & Evans (1998), only V measurements are quoted; (8) Hough et al. (1981); (9) Breger (1974), only $V$ measurements are quoted; (10) Vink et al. (2005); (11) Gledhill & Scarrott (1989); (12) Vrba et al. (1979), only $V$ measurements are quoted; (13) Hillenbrand et al. (1992); (14) Grinin et al. (1988), mean polarization on the photometric deep minima is quoted.

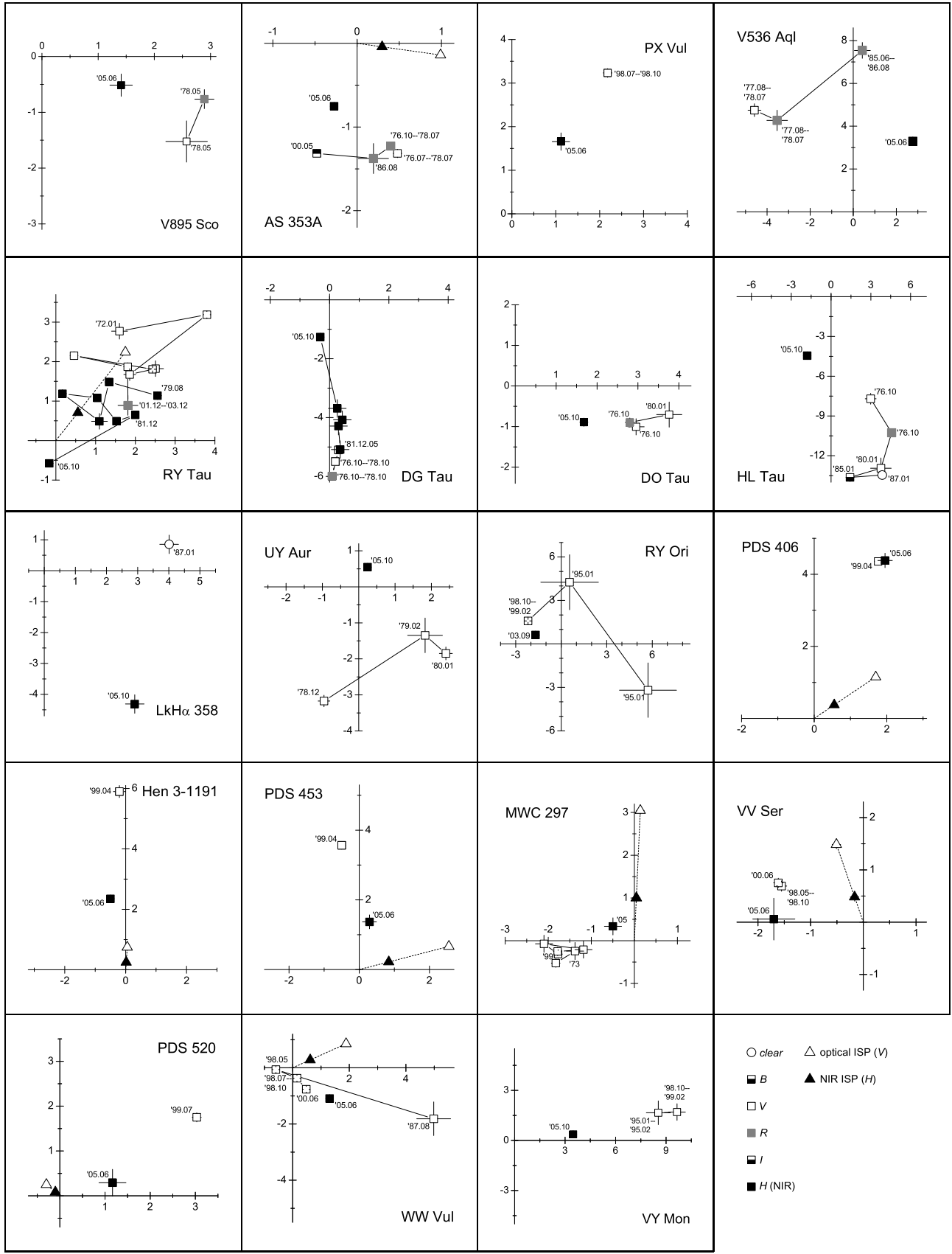

**Fig. 1.** $Q-U$ diagrams showing the temporal evolution for the objects of our sample. In each plot, the abscissa is the $Q$ parameter, and the ordinate, the $U$ parameter. Units are in percentage (%). The optical polarization data refer to Table 3. The measurements in *clear* filter are in white circles (only in HL Tau and LkH$\alpha$ 358), $V$ in white boxes, $R$ in grey boxes, $I$ in black-white boxes (only in AS 353A) and $B$ in white-black boxes (only in HL Tau). Our NIR polarization data (Table 3) and multiple epoch NIR data of RY Tau and DG Tau (Table 7) are shown in black boxes. The optical (and NIR, when available) temporal evolution is indicated by solid lines and labels. The direction of the foreground polarization, when available, is indicated by a dotted line. White and black triangles represent the optical and NIR foreground polarization vector, respectively.

**Table 4.** NIR linear polarization measurements.

| | | Observed Pol. | | | Foreground Pol.[a] | | | Intrinsic Pol. | |
|---|---|---|---|---|---|---|---|---|---|
| Object | Aper. ('') | $P_{obs}$ (%) | $PA_{obs}$ (°) | $P_{for}$ (%) | $PA_{for}$ (°) | N | $P_{int}$ (%) | $PA_{int}$ (°) |
| (1) | (2) | (3) | (4) | (5) | (6) | (7) | (8) | (9) |
| T Tauri | | | | | | | | |
| V895 Sco | 1.0 | 1.5 (0.2) | 170 (4) | – | – | – | – | – |
| AS 353A | 1.3 | 0.8 (0.04) | 125 (1) | 0.3 | 176 | | 0.9 (0.04) | 114 (1) |
| PX Vul | 5.0 | 2.0 (0.2) | 28 (3) | – | – | – | – | – |
| V536 Aql | 1.5 | 4.3 (0.2) | 25 (1) | – | – | – | – | – |
| RY Tau | 3.5 | 0.6 (0.02) | 127 (1) | 0.9 | 26 | | 1.5 (0.02) | 120 (0.4) |
| DG Tau | 2.5 | 1.3 (0.1) | 128 (2) | – | – | – | – | – |
| DO Tau | 6.0 | 1.9 (0.1) | 166 (1) | – | – | – | – | – |
| HL Tau | 0.5 | 4.8 (0.2) | 124 (1) | – | – | – | – | – |
| LkHα 358 | 1.0 | 5.2 (0.3) | 152 (2) | – | – | – | – | – |
| UY Aur | 6.0 | 0.6 (0.1) | 33 (5) | – | – | – | – | – |
| RY Ori | 5.5 | 1.8 (0.1) | 80 (1) | – | – | – | – | – |
| Herbig Ae/Be | | | | | | | | |
| PDS 406 | 1.5 | 4.8 (0.2) | 33 (1) | 0.7 (0.01) | 17 (0.4) | 348 | 4.2 (0.2) | 35 (1) |
| Hen 3-1191[b] | 1.6 | 2.4 (0.1) | 51 (1) | 0.2 (0.01) | 43 (1) | 546 | 2.1 (0.1) | 51 (1) |
| PDS 453 | 1.3 | 1.4 (0.2) | 39 (3) | 0.9 (0.02) | 7 (1) | 412 | 1.3 (0.2) | 58 (3) |
| MWC 297 | 4.0 | 0.6 (0.2) | 73 (10) | 1.0 (0.09) | 44 (3) | 4 | 0.9 (0.2) | 114 (8) |
| VV Ser | 3.5 | 1.7 (0.4) | 89 (6) | 0.5 (0.1) | 55 (4) | 5 | 1.6 (0.4) | 98 (7) |
| PDS 520 | 1.3 | 1.2 (0.3) | 7 (8) | 0.1 (0.03) | 70 (6) | 9 | 1.2 (0.3) | 4 (8) |
| WW Vul | 1.0 | 1.7 (0.1) | 160 (2) | 0.7 (0.01) | 12 (0.5) | 336 | 1.5 (0.1) | 148 (2) |
| VY Mon | 2.5 | 3.5 (0.2) | 3 (2) | – | – | – | – | – |

(a) from Rodrigues et al. (2009) except AS 353A, from Monin et al. (2006), and RY Tau, from Petrov et al. (1999).
(b) observed polarization is the average of two different nights (2005/jun/24 and 2005/jun/26).

**Table 5.** Correlations with position angles of jets and disks.

| Object | $PA_{jet}$ (°) | $PA_{disk}$ (°) | $\Delta_{jet/disk}$[a] (°) | $\Delta_{pol/disk}$[b] (°) | Disk opt. depth | Ref.[c] |
|---|---|---|---|---|---|---|
| T Tauri | | | | | | |
| AS 353A | 107 | – | – | 83 | thin | 1 |
| V536 Aql | 110 | – | – | 5 | thick | 2 |
| RY Tau | 115 | 27 (7) | 2 | 85, 87 | thin | 3,4 |
| DG Tau | 42 | ∼136 | 4 | 5, 9 | thick | 5,6 |
| DO Tau | 70 | 160 (9) | 0 | 6, 6 | thick | 7,8 |
| HL Tau | 51 | 125 (10) | 16 | 17, 1 | thick | 9,10 |
| LkHα 358 | 72 | – | – | 11 | thick | 11 |
| UY Aur | – | ∼135 | – | 78 | thin | 12 |
| Herbig Ae/Be | | | | | | |
| Hen 3-1191 | 48 | – | – | 87 | thin | 13 |
| PDS 453 | – | ∼45 | – | 13 | thick | 14 |
| MWC 297 | 164 | ∼100 | 26 | 40, 14 | thick | 15,16 |
| VV Ser | – | 15 (5) | – | 83 | thin | 17 |

(a) $\Delta_{jet/disk} = |PA_{jet} - PA_{disk} + 180 \times n - 90|$
(b) $\Delta_{pol/disk} = |PA_{pol} - PA_{axis} + 180 \times n - 90|$ when $PA_{axis} = PA_{jet}$; or $\Delta_{pol/disk} = 90 - \Delta_{pol/disk}$ when $PA_{axis} = PA_{disk}$; $PA_{pol}$ = $PA_{int}$ (when available) or $PA_{obs}$.
(c) references for $PA_{jet}$ and/or $PA_{disk}$: (1) Curiel et al. (1997); (2) Mundt & Eislöffel (1998); (3) Bastien & St-Onge (?); (4) Kitamura et al. (2002); (5) Lavalley et al. (1997); (6) Testi et al. (2002); (7) Hirt et al. (1994); (8) Koerner & Sargent (1995); (9) Mundt et al. (1990); (10) Wilner et al. (1996); (11) Moriarty-Schieven et al. (2006); (12) Potter et al. (2000); (13) Le Bertre et al. (1989); (14) Perrin (2006); (15) Drew et al. (1997); (16) Monnier et al. (2006); (17) Pontoppidan et al. (2007)

with data of different epochs, our conclusions can be biased by a long term intrinsic polarization variability and must be taken with care.

Nevertheless, within 25° in four objects (RY Tau, UY Aur, RY Ori and WW Vul) of our sample an polarization reversal seems to be present. The evidence of optically thin disks at NIR bands for RY Tau and UY Aur given in § 4.1 seems consistent with the polarization reversal model mentioned above. The indeterminacy of the disk optical depth of RY Ori and WW Vul prevents any conclusion about these two objects.

In particular, the $Q-U$ diagram for RY Tau shows a great variability already reported at optical and NIR bands. Our NIR polarization data on 2005/10 show a polarization reversal when comparing with the temporal evolution for the optical polarization (between 1972/01 and 2003/12). On the other hand, previous NIR polarization data that sample an interval of ∼2.5yr (between 1979/08 and 1981/12) are at the same quadrant ($Q > 0, U > 0$) than the historical optical variability. UY Aur was reported with a suspected optical polarization variability (Bastien 1985). This object shows a polarization reversal if we compare the optical polarization on 1978/12/01 with our NIR polarization. However, the feature is not evident if the comparison is done with the optical data between 1979/02 and 1980/01. In RY Ori the detection of the polarization reversal is marginal. The feature is evident only between the first of the two optical data on 1995/01/30 and our NIR data. Finally, in WW Vul polarization reversal is evident between the optical data on 1998/05 and our NIR data but fails if the comparison is done with the optical data on 1998/07−10 and 2000/06.

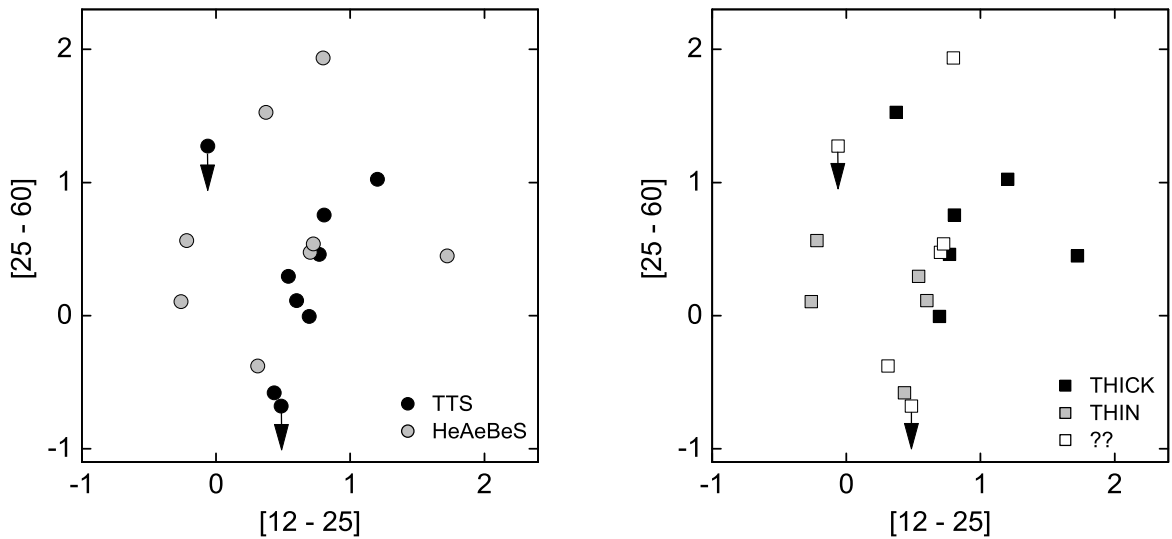

**Fig. 4.** *IRAS* colour diagram for our sample. (*left*) The *locus* for the TTS (black dots) and the HeAeBeS (grey dots) is shown. (*right*) The optically thick disks (black squares) in our sample seems to ocuppy a different *locus* than the optically thin disks (grey squares). Objects with an undetermined disk optical depth also are shown (white squares).

It is interesting to note that RY Tau, RY Ori, and WW Vul with evidence of polarization reversal present the UXOR phenomenon (Grinin 1994; Yudin & Evans 1998; Petrov et al. 1999; Oudmaijer et al. 2001). Nevertheless, VY Mon which also shows the UXOR behaviour (Oudmaijer et al. 2001) does not show the polarization reversal in our comparison (Figure 1). The UXOR phenomenon is associated with an increase in polarization simultaneous to an abrupt decrease in the optical brightness of the central object (Grinin et al. 1994) in non-periodic variables with deep Algol-type minima. This feature is explained by the effect of dust clumps rotating around the central star in a disk-like configuration. Clumps in our line of sight will absorb the light of the central star increasing the proportion of scattered light and consequently the polarization. When the line of sight is free of clumps, the star will become bright again and the contribution of the scattered light will be reduced along with the polarization. In the optical bands, polarization PA flips are common in stars with the UXOR phenomenon. For example, UX Ori (the prototype for the UXOR phenomenon) shows after the minimum a rapid increase of the polarization level followed by a gradual polarization PA flip of ∼90° (Grinin 1994). A similar behavior also was reported for WW Vul with a polarization PA flip of ∼60° (Grinin et al. 1988) and recently for CO Ori (Rostopchina et al. 2007, PA flip ∼90°).

In RY Tau the historical NIR polarization PA variability would imply significant changes between an optically thick disk between 1979/08−1981/12 (Hough et al. 1981; Moneti et al. 1984) and an optically thin disk in our NIR data. These changes are probably associated with the screening effect by clumps following the UXOR behavior.

### 4.3. IRAS colours and spectral index

In order to determine if any correlations exist between the disk optical depth computed in § 4.1 with the *IRAS* colours, we plotted in Fig. 4 the *IRAS* colour diagram for our sample. The far-IR colours were calculated using $[\lambda_i - \lambda_j] = 2.5\log[F(\lambda_j)/F(\lambda_i)]$, where $F(\lambda)$ are the *IRAS* fluxes (Table 6). Only V895 Sco and LkHα 358 have no *IRAS* counterpart. The HeAeBeS and TTs are spanned in an approximately homogeneous range on the diagram (Fig. 4, left). However, when the positions on the diagram are classified by the disk optical depth is evident a segregation between the two subsamples (Fig. 4, right). The optically thick disks present more far-IR emission than the optically thin ones and an evolutionary segregation is evident. For completeness, we also plotted the objects with undetermined disk optical depth. If the segregation is true, the objects with more far-IR emission are expected to have optically thick disks.

Another way to check the last result is computing the spectral index for the objects of our sample with a well determined disk optical depth (Fig. 5). We used the classification by Lada (1987) with $\alpha(K, 25) = \log[\lambda_{25}F_K/(\lambda_K F_{25})]/\log(\lambda_K/\lambda_{25})$ representing the spectral index between the 2MASS $K$ band and the

**Table 6.** Additional information.

| Object | IRAS name | [12-25] | [25-60][a] | $\alpha(K,25)$ |
|---|---|---|---|---|
| T Tauri | | | | |
| AS 353A | 19181+1056 | 0.54 | 0.30 | -0.06 |
| PX Vul | 19245+2347 | 0.49 | -0.68 | -0.74 |
| V536 Aql | 19365+1023 | 0.69 | -0.01 | -0.15 |
| RY Tau | 04188+2819 | 0.43 | -0.58 | -0.29 |
| DG Tau | 04240+2559 | 0.81 | 0.76 | 0.19 |
| DO Tau | 04353+2604 | 0.77 | 0.46 | -0.35 |
| HL Tau | 04287+1807 | 1.20 | 1.02 | 0.53 |
| UY Aur | 04486+3042 | 0.60 | 0.11 | -0.16 |
| RY Ori | 05296−0251 | -0.06 | 1.27 | -0.68 |
| Herbig Ae/Be | | | | |
| PDS 406 | 16017−3936 | 0.80 | 1.94 | 0.26 |
| Hen 3-1191 | 16235−4832 | -0.26 | 0.11 | -0.11 |
| PDS 453 | 17178−2600 | 1.72 | 0.45 | 0.10 |
| MWC 297 | 18250−0351 | 0.37 | 1.53 | -0.30 |
| VV Ser | 18262+0006 | -0.22 | 0.56 | -0.74 |
| PDS 520 | 18275+0040 | 0.70 | 0.47 | -0.29 |
| WW Vul | 19238+2106 | 0.31 | -0.38 | -0.59 |
| VY Mon | 06283+1028 | 0.72 | 0.54 | 0.12 |

(a) Upper limits for PX Vul and RY Ori.

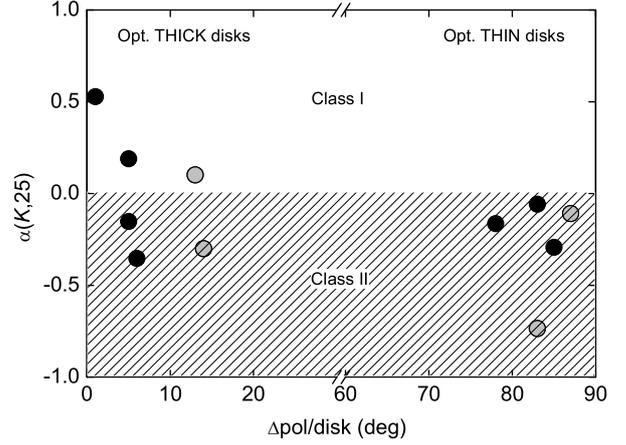

**Fig. 5.** Spectral index $\alpha(K,25)$ for the YSOs in our sample with $\Delta_{\rm pol/disk}$ parameter determined. Class I and II sources are equally distributed in optically thick disks. Nevertheless, the optically thin disks are only associated to class II sources. The TTS are in black dots and the HeAeBeS in grey dots.

$25\mu m$ IRAS flux (see Table 6). For class I sources $0 < \alpha \le 3$ and for class II sources $-2 \le \alpha \le 0$. Clearly, all the optically thin disks are associated to class II sources. On the other hand, the optically thick disks appear equally distributed between the two evolutionary classes. The link between the class II objects and the optically thin disks reinforces the idea of an evolutionary segregation for the YSOs disks. If this is correct, we have found evidence that the disks in YSOs tend to be optically thin when they are older. This picture is consistent with the final phases of the disk evolution in the standard star formation model (Shu et al. 1987).

## 5. Comments on individual objects

### 5.1. T Tauri stars

#### 5.1.1. AS 353A

AS 353A is a TTS member of a binary system with a separation of 5″.7 (Reipurth & Zinnecker 1993). The aperture radius used in our measurement is 1″.3 that avoids contamination by the companion. The outflow axis direction, PA =107° (or PA =287°), is well determined by the relative positions of its associated HH objects (HH 32A-D, Hartigan et al. 1986) and proper motions (Curiel et al. 1997). Our NIR polarization PA (125°±1°) is approximately parallel to the outflow axis. If we consider the ISP correction (PA =114°±1°), the correlation is even better (Fig. 6). This is consistent with the NIR polarization being produced in an optically thin disk.

#### 5.1.2. V536 Aql

V536 Aql is a classical TTS (CTTS). Optical polarimetric variability was reported by Ménard & Bastien (1992) with a range of polarization ~[5.5−8.5]% and PA ~[40−70]° measured during about 9 yrs. Subarcsecond NIR imaging (Ageorges et al. 1994) for V536 Aql detected a binary system with a PA axis ~17° and an angular distance of ~0.5″. Evidence of the outflow/jet axis was found by Hirth et al. (1997) using long-slit spectra of forbidden emission lines (PA =90°±20°); and by Mundt & Eislöffel (1998) using [SII]−continuum imaging (PA ~110°). Our measurements do not resolve the system but the NIR polarization PA (25°±1°) is approximately parallel to the binary PA axis and also perpendicular to the bipolar outflow (Fig. 7). Probably, the NIR polarization is produced in an optically thick dusty (and circumbinary?) disk.

#### 5.1.3. RY Tau

An estimate of the disk PA is found in Kitamura et al. (2002, PA =27°±7°) using 2mm thermal dust continuum emission, and Koerner & Sargent (1995, PA =48°±5°) using CO(2→1) emission. Considering the consistent disk evidence from CO and 2mm emission, the NIR polarization PA (127°±1°) is approximately perpendicular to the disk PA. The correlation is even better if we consider the correction by ISP (PA$_{int}$ =120°±0°.4, Table 3). Recently, St-Onge & Bastien (2008) using H$\alpha$−continuum Gemini imaging reported a bipolar jet from RY Tau with PA =115° and approximately parallel to our NIR polarization PA (Fig. 8). This seems to confirm that an optically thin disk is more likely in RY Tau. Nevertheless, previous NIR polarization measurements have been reported with polarization PA different to our value (see Table 7). This is a clear indication of a great variability of this source also at the NIR bands. As we noted in § 4.2 probably the UXOR phenomenon can explain the large variations on the polarization PA.

#### 5.1.4. DG Tau

The NIR polarization PA (128°±2°) is roughly perpendicular to the jet direction (PA =222°) obtained from [OI]−continuum maps (Lavalley et al. 1997). Evidence of a Keplerian disk rotating perpendicular to the jet is found by Testi et al. (2002) using $^{13}$CO wing emission. Therefore, the NIR polarization PA is also parallel to the disk as expected in an optically thick dusty disk (Fig. 9). It is interesting to note that previous published polarization data in errors in parenthesis.
References: (1) Hough et al. (1981); (2) Moneti et al. (1984); (3) Lucas et al. (2004) with aperture radius of 0.2″.

**Table 7.** NIR polarization in $H$ band for RY Tau, DG Tau and, HL Tau from the literature.

| Object | Date | $P$ (%) | PA (°) | Ref. |
|---|---|---|---|---|
| RY Tau | 1979/aug/12 | 2.8 (0.1) | 12 (1) | 1 |
|  | 1979/oct/13 | 2.0 (0.1) | 24 (1) | 1 |
|  | 1979/nov/12 | 1.2 (0.2) | 12 (2) | 1 |
|  | 1980/feb/27 | 1.2 (0.1) | 41 (3) | 1 |
|  | 1981/dec/05 | 1.5 | 23 | 2 |
|  | 1981/dec/06 | 1.6 | 9 | 2 |
|  | 1981/dec/07 | 2.1 | 9 | 2 |
| DG Tau | 1981/dec/05 | 5.1 (0.3) | 137 | 2 |
|  | 1981/dec/06 | 4.3 (0.3) | 137 | 2 |
|  | 1981/dec/07 | 4.1 (0.3) | 138 | 2 |
|  | 1982/oct/04 | 3.7 (0.3) | 137 | 2 |
| HL Tau | 2000/dec/28 | 3.7 (0.3) | 77 | 3 |

$H$ band for DG Tau (see Table 7) show a higher polarization level but with the PA consistent with ours within 10°. Therefore, the polarimetric variability seems not connected with geometric changes in the material distribution.

### 5.1.5. DO Tau

The NIR polarization PA (166°±1°) is consistent with the optical value (PA =170°, Bastien 1982). Hirth et al. (1994) detected the optical jet direction (PA =70°) using [SII]. Koerner & Sargent (1995) determined the disk PA (160°±9°) from CO(2→1) emission. The optical jet and the CO emission seem consistent with the NIR polarization PA being produced in an optically thick dusty disk (Fig. 10). However, Kitamura et al. (2002) estimated a different (and perpendicular) disk PA (67°±9°) from dust thermal emission at 2mm. Probably, this discrepancy can be explained by the poorly resolved dust emission. The difference between the longer and smaller disk axes, in that work, is lower than 20%.

### 5.1.6. HL Tau

HL Tau is a well studied T Tauri star. Mundt et al. (1990) determined the jet direction (PA =51°) using [SII]−continuum imaging. Several works using different techniques determined a consistent disk PA around 125° (Lay et al. 1994; Mundy et al. 1996; Wilner et al. 1996; Close et al. 1997; **Murakawa et al. 2008**). Different disk PA are found in Sargent & Beckwith (1991, PA =160°), Stapelfeldt et al. (1995, PA =145°), Kitamura et al. (2002, PA =144°±2°), Lucas et al. (2004, PA =136°±8°), and Rodmann et al. (2006, PA =170°±30°). The jet and disk evidence along with our NIR polarization PA (124°±1°) are consistent with an optically thick dusty disk in HL Tau (Fig. 11).

### 5.1.7. LkHα 358

Recently, Moriarty-Schieven et al. (2006) have suggested that LkHα 358 has a jet with PA =72°. This jet direction is rigorously perpendicular to the NIR polarization PA (152°±2°). Therefore, this is consistent with an optically thick disk in LkHα 358 (Fig. 12).

### 5.1.8. UY Aur

UY Aur is a binary system with a separation of ∼0″.9 and a binary axis with PA =225° (Leinert et al. 1993). NIR polarization maps at high angular resolution (Potter et al. 2000) yielded a PA for the circumbinary disk of 135°. The large circumbinary disk begins ∼2″ from the center of the binary system and goes up to ∼5″. We selected a radius aperture of 6″ to properly cover the circumbinary disk. Our NIR polarization PA is 33°±5° approximately perpendicular to the disk PA. It seems consistent with a probably optically thin circumbinary disk in UY Aur.

### 5.2. Herbig Ae/Be stars

### 5.2.1. PDS 406

PDS 406 is a Herbig Ae star, still embedded in its parental cloud, as suggested by the significant visual extinction ($A_V$=1, Chen et al. 1997) and the high NIR foreground polarization (0.7%, Table 3) toward this object. As we noted in § 3.3, PDS 406 is interesting because it is the object in our sample with the largest $P_{\rm int}$ (4.2%). Viera et al. (2003) detected an Hα double peak that may indicate a rotating disk. Low resolution (30″) maps from $^{12}$CO(J=2→1) (Tachihara et al. 1996) did not detect molecular outflows in PDS 406. High-resolution measurements are needed for this object.

### 5.2.2. Hen 3-1191

The nature of Hen 3-1191 is not clear. It was classified as a proto-planetary nebula (PPN) by Le Bertre et al. (1989). However, De Winter et al. (1994) concluded that Hen 3-1191 is likely a pre-main sequence star. Imaging in $R$ band (Le Bertre et al. 1989) shows clearly a bipolar structure with an axis oriented NW-SE (PA ∼48°). The NIR polarization PA (51°±1°, corrected by ISP) is rigorously parallel to the bipolar axis (Fig. 13). This is consistent with scattering in an optically thin disk in Hen 3-1191. Recently, Lachaume et al. (2007) resolved Hen 3-1191 at $N$ band (8-13$\mu$m) with VLTI/MIDI and found a disk PA of 1°±4°. The disagreement with our measurement may be related with the high angular resolution (24-36mas) from Lachaume et al. (2007) that maps the more interior disk region.

### 5.2.3. PDS 453

PDS 453 was classified as a possible PPN by Kohoutek (1997) – see also García-Lario et al. (1997). On the other hand, Vieira et al. (2003) classify PDS 453 as a Herbig Ae/Be candidate but without a clear association to star-forming regions (possibly LDN 1767 and/or 1773). Recently, Perrin (2006) at NIR bands detected a possible polarization disk with PA ∼45° explained by multiple scattering. This PA is approximately parallel to our NIR polarization PA (58°±3°, corrected by ISP) and consistent with an optically thick disk.

### 5.2.4. MWC 297

MWC 297 is a Herbig Be star with a well-defined spectral type (B1.5Ve, Drew et al. 1997). Monnier et al. (2006) using NIR interferometry estimated an elongation of the inner disk with a PA ∼100°. This disk PA is approximately parallel to our NIR polarization PA (114°±8°, corrected by ISP) and consistent with the NIR polarization produced by multiple scattering (Fig. 14). This seems to confirm previous evidence of an optically thick disk by fitting the SED at mm/submm (Mannings 1994) and modeling NIR interferometry data (Malbet et al. 2007). Consistent with this scenario, the radio emission associated with this source (PA =164°, Drew et al. 1997) is marginally perpendicular with respect to our NIR polarization PA.

### 5.2.5. VV Ser

VV Ser is a Herbig Ae star. Eisner at al. (2004) estimated a disk PA in the 165-173° range using NIR interferometry. Recently, Pontoppidan et al. (2007) studied and modeled Spitzer data of IR continuum and PAH emission for this object with a disk shadow PA shadow of 15°±5°. Our NIR polarization PA (98°±7°, corrected by ISP) is approximately perpendicular to the disk PA from Eisner et al. (2004) and Pontoppidan et al. (2007). Therefore, an optically thin disk is more likely in VV Ser (Fig. 15).

### 5.2.6. VY Mon

VY Mon is an eruptive, highly reddened ($A_V$ ∼7.4mag), Algol-like variable Herbig Ae/Be star. VY Mon is associated with the reflection nebula IC 446 (1.5′ north) and a Bok globule (1.5′ south). There is not clear evidence for a jet/disk in VY Mon. However, evidence of asymmetry was found by modeling the FIR emission (Casey & Harper 1990). Henning et al. (1998) also found $K$-band and milimeter emissions slightly extended to the south of VY Mon. Our NIR polarization PA (3°±2°) is parallel to the axis defined by IC 446 and the Bok globule and also by the millimeter emission. However, it is not clear whether the dust emission (from de Bok globule) and the reflection nebula are tracing the same direction as the circumstellar disk.

## 6. Conclusions

We present NIR polarimetric measurements of a sample of YSOs. Except for RY Tau, DG Tau, and HL Tau, the remaining sixteen objects have their NIR polarization published by the first time.

Eight TTS and four HeAeBeS of our sample have its jet/disk orientation determined by other means. It enables us to correlate the NIR polarization PA with the disk PA and hence to determine the optical depth of the disk. Three TTS present optically thin disks (AS 353A, RY Tau, and UY Aur) and five optically thick disks (V536 Aql, DG Tau, DO Tau HL Tau, and LkHα 358). On the other hand, two HeAeBeS show evidence of optically thin disk (Hen 3-1191 and VV Ser) and two of optically thick disks (PDS 453 and MWC 297). Therefore, considering the objects with correlations available, we found that ∼60% (7 of 12 objects) have optically thick disks at NIR bands. Our statistics is limited but our sample seems to confirm the trend of optically thick disks in YSOs (Ménard & Bastien 1992; Maheswar et al. 2002) obtained at the optical bands.

Marginal evidence for polarization reversal between the optical and NIR bands is found in four objects (RY Tau, UY Aur, RY Ori, and WW Vul). The presence of optically thin disks on the epoch of our observations for RY Tau and UY Aur given by the correlation between the NIR polarization and the jet/disk PA seems consistent with the polarization reversal model. Interestingly, RY Tau, RY Ori, and WW Vul, which show evidence of polarization reversal, present the UXOR behavior. In particular, RY Tau shows historical changes on the disk optical depth at NIR bands that we believe influenced by the presence of eventual dust clumps on the line of sight associated to the UXOR behavior. Probably, RY Ori and WW Vul must have a similar behavior but more data are necessary to better constrain this issue.

Finally, we show a correlation between the *IRAS* colours and the disk optical depth. The optically thick disks are located in a place on the colour diagram that corresponds to larger values of far-IR emission. This can be interpreted as an evolutionary segregation. It is corroborated by the fact that only objects classified as Class II sources by their spectral indexes present optically thin disks. Therefore, polarimetric measurements at NIR bands along with an independent evidence of the geometry of the PA disk can potentially help to constrain the evolutionary stage of YSOs.

*Acknowledgements.* The authors wish to thank the anonymous referee for his/her careful reading. His/her several comments and suggestions helped to improve the paper. A. P. thanks FAPESP (grant 02/12880-0) and CNPq (DTI grant 382.585/07-03 associated with the PCI/MCT/ON program). A. P. is also grateful to ALFA/LENAC project for financial support at ICE/Spain. A. M. M. is thankful to FAPESP and CNPq for financial support. J. M. G. is supported by MEC grant AYA2005−08523−C03 (including FEDER funds) and AGAUR grant 2005SGR 00489. Polarimetry at IAG-USP is supported by FAPESP grant 01/12589−1. This research has made use of the VizieR catalogue access tools operated at CDS, Strasbourg, France.

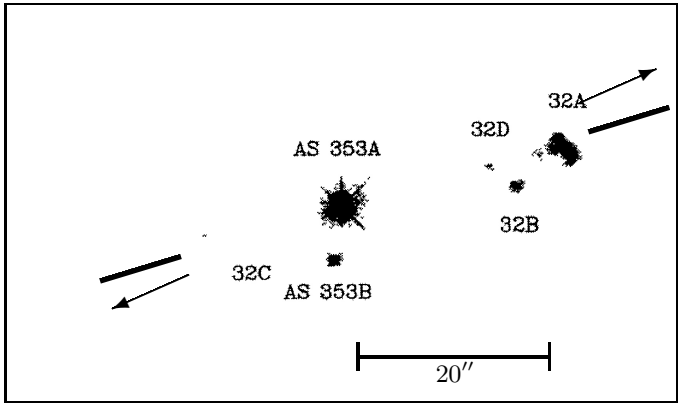

**Fig. 6.** Hα image of AS 353A adapted from Curiel et al. (1997). The HH objects associated are shown along with the outflow axis defined by them (thick lines). The PA from NIR polarization (corrected by ISP) is also indicated (black arrows). North is top and East is left.

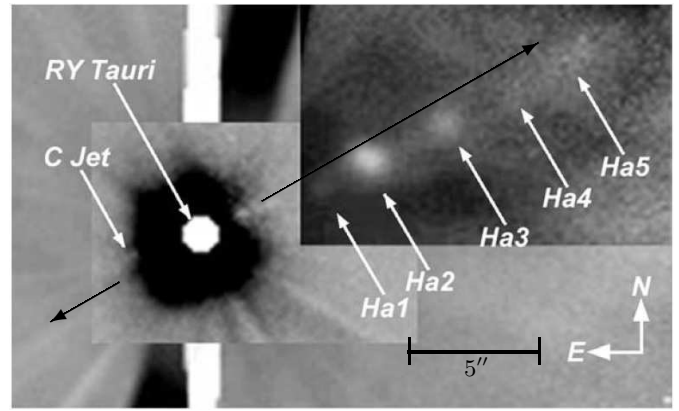

**Fig. 8.** Hα−continuum image of RY Tau adapted from St-Onge & Bastien (2008). The knots (Ha1 to Ha5) indicate the jet axis. The PA from NIR polarization (corrected by ISP) is also indicated (black arrows).

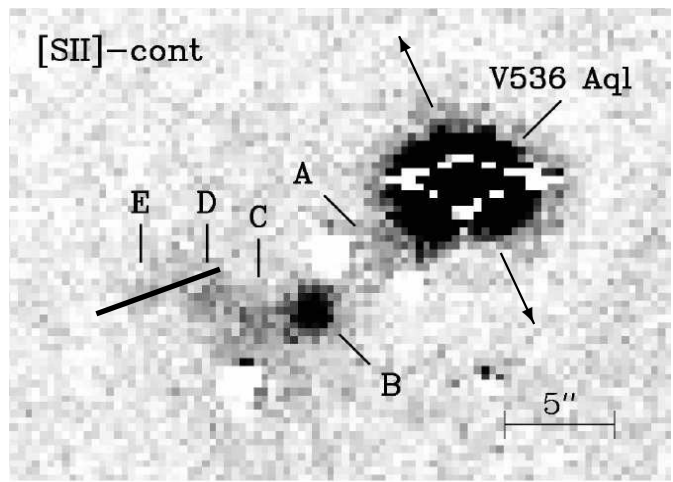

**Fig. 7.** [SII]-continuum image of V536 Aql adapted from Mundt & Eislöffel (1998). The jet axis is indicated (thick line) along with the PA from NIR polarization (arrows). North is top and East is left.

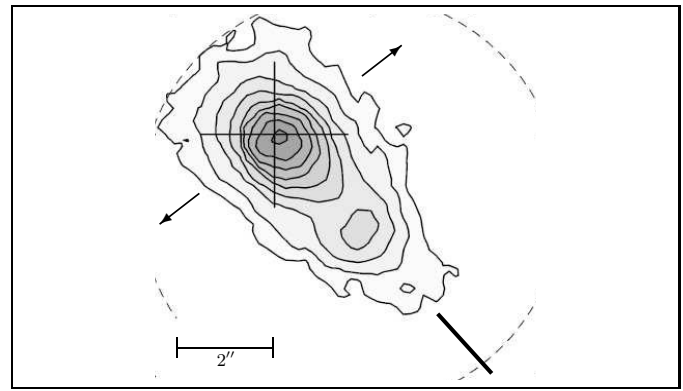

**Fig. 9.** [OI]-continuum image of DG Tau adapted from Lavalley et al. (1997). The jet axis is indicated (thick line) along with the PA from NIR polarization (arrows). North is top and East is left.

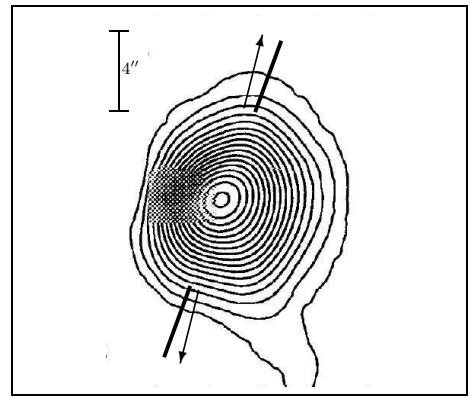

**Fig. 10.** CO emission map for DO Tau adapted from Korner & Sargent (1995). The disk PA from CO is shown (thick line) along with the PA from NIR polarization (arrows). North is top and East is left.

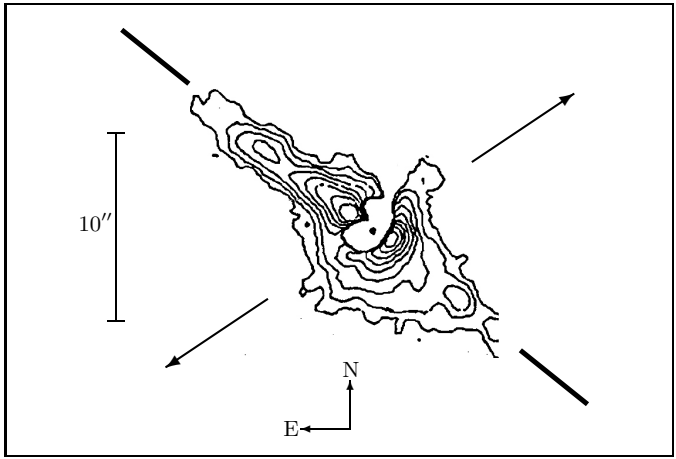

**Fig. 11.** [SII]-$r_N$ image of HL Tau adapted from Mundt et al. (1990) indicating the jets associated (thick line). The PA from NIR polarization is also indicated (arrows).

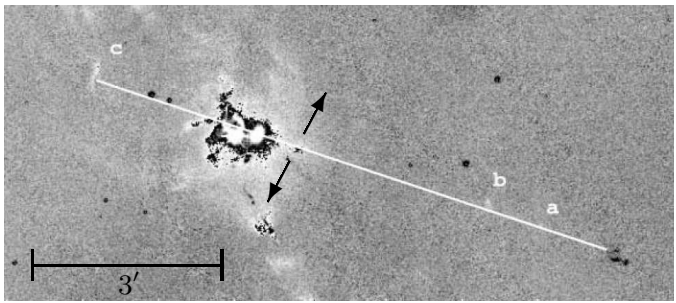

**Fig. 12.** Jet axis and knots associated (a, b and c) for LkH$\alpha$ 358 adapted from Moriarty-Schieven et al. (2006, white line). The PA from NIR polarization (arrows) is also indicated. North is top and East is left.

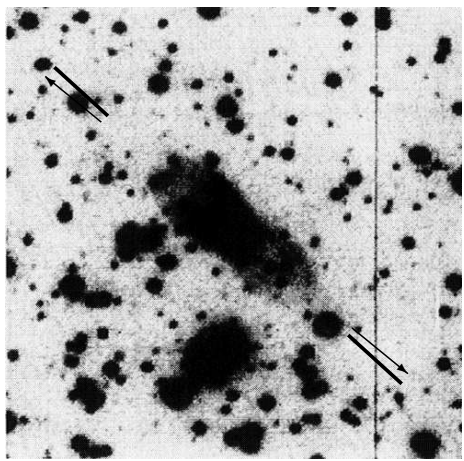

**Fig. 13.** $R$ frame of Hen 3-1191 from Le Bertre et al. (1989) showing the bipolar structure (axis in thick line). The field is $2' \times 2'$. The PA from NIR polarization is shown (arrows). North is top and East is left.

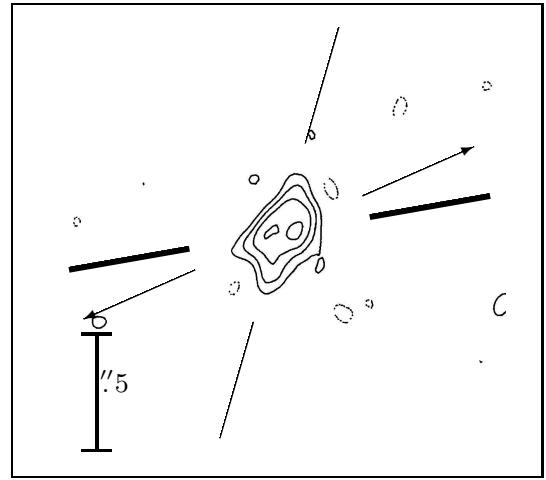

**Fig. 14.** MERLIN 5Ghz radio map of MWC 297 adapted from Drew et al. (1997). The outflowing axis at 5Ghz is indicated (thin line). The IOTA interferometer disk PA at NIR from Monnier et al. (2006) also is shown (thick line) along with the PA from NIR polarization corrected by ISP (arrows). North is top and East is left.

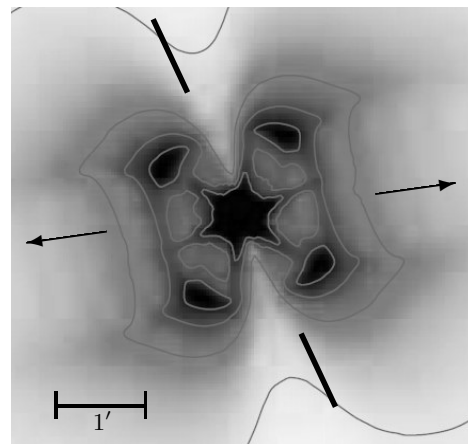

**Fig. 15.** IRAC 8$\mu$m model of VV Ser adapted from Pontoppidan et al. (2007) with the disk PA indicated (thick line). The PA from NIR polarization corrected by ISP is also shown (arrows). North is top and East is left.